\documentclass[aps,pre,reprint,longbibliography,floatfix,groupedaddress]{revtex4-1}
\usepackage{graphicx}
\usepackage{hyperref}
\usepackage{dcolumn}
\usepackage{color}
\usepackage{amsmath,amssymb,amsfonts}
\graphicspath{{figs/}}

\begin{document}

\title{Percolation of sticks: effect of stick alignment and length dispersity}

\author{Yuri~Yu.~Tarasevich}
\email[Corresponding author: ]{tarasevich@asu.edu.ru}

\author{Andrei~V.~Eserkepov}
\email{dantealigjery49@gmail.com}

\affiliation{Laboratory of Mathematical Modeling, Astrakhan State University, Astrakhan, Russia, 414056}


\date{\today}

\begin{abstract}
Using Monte Carlo simulation, we studied the percolation of sticks, i.e. zero-width rods, on a plane paying  special attention to the effects of stick alignment and their length dispersity. The stick lengths were distributed in accordance with  log-normal distributions, providing a constant mean length with different widths of distribution.  Scaling analysis was performed to obtain the percolation thresholds in the thermodynamic limits for all values of the parameters. Greater alignment of the sticks led to increases in  the percolation threshold while an increase in length dispersity decreased the percolation threshold. A fitting formula has been proposed for the dependency of the percolation threshold both on stick alignment and on length dispersity.
\end{abstract}


\maketitle

\section{Introduction}\label{sec:intro}
Percolation, i.e., the occurrence of a connected subset (a cluster) within a disordered medium which spans its opposite borders, has attracted the attention of the scientific community over several  decades~\cite{Stauffer,Sahimi1994,BollobasRiordan2006,Grimmett1999,Kesten1982}. Nowadays, two-dimensional (2D) systems such as transparent electrodes present examples of where highly conductive particles, e.g., nano-wires (NWs),  nano-tubes (NTs), and nano-rods (NRs), form a random resistor network (RRN) inside a poorly  conductive host matrix (substrate)~\cite{Mutiso2015PPS,Kumar2016JAP,Kumar2017JAP}. The appearance of a percolation cluster in this kind of systems drastically changes their physical properties and  is associated with an insulator-to-conductor phase transition. Length dispersity is common for NWs, NTs, and NRs~\cite{Kyrylyuk2008PNAS,Gnanasekaran2014,Borchert2015N,Majidian2017SR}. These works evidenced that the length distributions of NWs, NTs, and NRs are close to representing log-normal distributions. Furthermore, the alignment of such elongated objects may be produced in a variety of different ways~\cite{Chen2004JVSTB,Du2005PRB,Park2006JPSB,Behnam2007JAP,Yang2011NML}. Both length dispersity and alignment affect the electrical conductivity of the samples~\cite{Hicks2009PRE,Ackermann2016SR}.

Study of the percolation of rod-like particles or sticks in 2D systems and its connection with the electrical conductivity has a long history~\cite{Pike1974PRB,Balberg1982SSC,Balberg1983PRL}. At present, the best known value of the percolation threshold in 2D system of randomly oriented and placed zero-width sticks of equal length is 5.637\,285\,8(6) sticks per unit area~\cite{Mertens2012PRE}. The number of objects per unit area is also known as the number density.

A computer study of the percolation threshold in a two-dimensional anisotropic system of conducting sticks has been performed~\cite{Balberg1983PRB}. Here, two kinds of angle distributions were taken into consideration, viz., uniform distribution within an interval,
\begin{equation}\label{eq:uniformangdistr}
-\theta_m \leqslant \theta \leqslant \theta_m,
\end{equation}
and the normal distribution; log-normal distribution of lengths was assumed. An analytical relationship  between the critical density of sticks and anisotropy has been proposed.  This relation predicts that the percolation threshold will increase with increasing anisotropy from its isotropic value. Obviously, in a system of completely aligned, i.e., parallel, sticks, no percolation can occur.

The conductivity of stick percolation clusters with anisotropic alignments has been studied by means of computer simulation and finite-size scaling analysis~\cite{Yook2012JKPS}. The angular distribution of the sticks corresponded to~\eqref{eq:uniformangdistr}. The critical number density, $n_c$, does not vary much for $\theta_m \in (5\pi/18,\pi/2]$ while it changes rapidly as
\begin{equation}\label{eq:ncfit}
  n_c \sim \theta_m^{-0.9}
\end{equation}
for $\theta_m < 5\pi/18$. The percolation threshold (critical number density) increases rapidly as the anisotropy is increased.

The finite continuum percolation of rectangles with different aspect ratios has been studied using their angular distribution~\cite{Klatt2017JSMTE}
$$
f_\theta(\theta) = \frac{\Gamma\left( \frac{\alpha}{2} + 1 \right)}{\sqrt{\pi} \Gamma \left( \frac{\alpha + 1}{2}\right)} \cos^\alpha \theta, \quad \theta \in \left[-\frac{\pi}{2}, \frac{\pi}{2}\right).
$$
The parameter $\alpha$ controls the degree of anisotropy of the system. $\alpha = 0$ corresponds to a uniform distribution $f_\theta(\theta) = \pi^{-1}$ and hence to an isotropic system. The larger the value of $\alpha$ the stronger the anisotropy.  $\alpha= \infty$ corresponds to a full alignment of the sticks along the $x$-axis.

Furthermore, the effect of the length dispersity of sticks on the percolation threshold has also been studied in several works. For instance, sticks with log-normal distributions of lengths were considered in~\cite{Balberg1983PRB}. The effects of length distribution, angular anisotropy, and wire curvature have been investigated both numerically and experimentally~\cite{Langley2018NH}. Each of these quantities was assumed to be normally distributed.  The percolation threshold decreases as either the length or the angle dispersity increases. Furthermore, the cooperative influence of both effects, simultaneously, on the percolation threshold may be of special interest.

Percolation in systems of aligned rods with different aspect ratios has been simulated~\cite{Chatterjee2014JCP}. Both systems of rods of equal length and systems consisting of mixtures of short and long rods were considered. Alignment of the rods led to increases in the percolation threshold. For mixtures of long and short rods, nomonotonic dependence of the percolation threshold on the fraction of short rods was demonstrated.

Numerical simulations of stick percolation have been performed~\cite{Mietta2014JCP} with uniform angular distributions of the sticks within a given interval as well as with normal distributions, while the stick lengths corresponded to a log-normal distribution. The probabilities of percolation were presented for different values of the parameters.

The goal of the present work was to obtain more accurate values for the dependencies of the percolation thresholds on anisotropy and length dispersity. The rest of the paper is constructed as follows. In Section~\ref{sec:methods}, the technical details of the simulations and calculations are described. Section~\ref{sec:results} presents our main findings. Section~\ref{sec:concl} summarizes the main results.

\section{Methods}\label{sec:methods}
\subsection{Preparation of the film samples}
Zero-width (widthless) sticks were deposited randomly and uniformly with given anisotropy onto a substrate  of size $L \times L$ having periodic boundary conditions (PBCs), i.e., onto a torus. Intersections of the particles were allowed. The length of the particles, $l$, varied according to a log-normal distribution with the probability density function (PDF)
\begin{equation}\label{eq:lognorm}
  f_l(l)=\frac{1}{l\sigma_l\sqrt{2\pi }}\exp \left( -\frac{\left( \ln l-\mu_l \right)^2}{2\sigma_l^2} \right).
\end{equation}
The mean, $\langle l \rangle$, and the standard deviation, $\mathrm{SD} (l)$, are connected with the parameters of the log-normal distribution, $\mu_l$, $\sigma_l$,  as follows
\begin{equation}\label{eq:mean}
  \langle l \rangle = \exp\left(\mu_l+\frac{\sigma_l^2}{2}\right),
\end{equation}
\begin{equation}\label{eq:var}
  \mathrm{SD} (l)^2 = \left(\exp\left(\sigma_l^2\right)-1\right) \exp \left(2\mu_l+\sigma_l^2\right).
\end{equation}
A change of any parameter affects both the mean and the standard deviation. To avoid a superposition of different effects, the mean was set as a constant during the simulations. In this case, we could extract and study the individual effect of the length dispersity. All our computations were performed for $\langle l \rangle = 1.$
For this particular value of the mean, the parameters of the log-normal distribution are
$$
\mu_l = -\frac{\sigma_l^2}{2}, \quad  \sigma_l^2 = \ln \left(\mathrm{SD} (l)^2 + 1\right).
$$

The anisotropy of the system is characterized by  the order parameter (see, e.g.,~\cite{Frenkel1985PRA})
\begin{equation}\label{eq:s}
s=N^{-1}\sum_{i=1}^N \cos 2\theta_i,
\end{equation}
where $\theta_i$ is the angle between the axis of the $i$-th stick and the horizontal axis $x$, and $N$ is the total number of sticks in the system. Since for a uniform angular distribution within a symmetric interval~\eqref{eq:uniformangdistr}
$$
s = \frac{\sin 2\theta_m}{2\theta_m},
$$
relation~\eqref{eq:ncfit} can be rewritten as
\begin{equation}\label{eq:asymp}
n_c \sim (1 - s)^{-0.45}.
\end{equation}
Furthermore, the macroscopic anisotropy
\begin{equation}\label{eq:A}
  A = \frac{\langle l | \cos \theta|\rangle}{\langle l | \sin \theta|\rangle}
\end{equation}
was used~\cite{Balberg1983PRB,Mietta2014JCP} to characterize the anisotropy of systems with length dispersity. Here, $\langle \cdot \rangle$ denotes the mean value.

In our simulations, the angles were distributed according to a normal distribution~\cite{Tarasevich2018PREa}
\begin{equation}\label{eq:angledistrib}
  f_\theta(\theta) = \frac{1}{\sqrt{- \pi \ln s}} \exp\left(\frac{\theta^2}{\ln s} \right).
\end{equation}

For each sample, a sequence of random positions (two coordinates for each stick), orientations, and lengths was generated. This sequence was used to produce a film with the desired number density of sticks, $n$,
\begin{equation}\label{eq:numdens}
  n = \frac{N}{L^2}.
\end{equation}
Since support of the log-normal distribution is $l \in (0, \infty)$, the probability that $l>L$ is finite, although very small. All sticks with $l>L$ were rejected for deposition and excluded from the sequence.

We performed our simulations for different values of the order parameter and length dispersity, viz., $s=0, 0.1, \dots, 0.9$, and  $\mathrm{SD} = 0, 0.5, 1.0$.

\subsection{Estimation of the percolation threshold}

\begin{figure}[!htb]
  \centering
  \includegraphics[width=0.9\columnwidth]{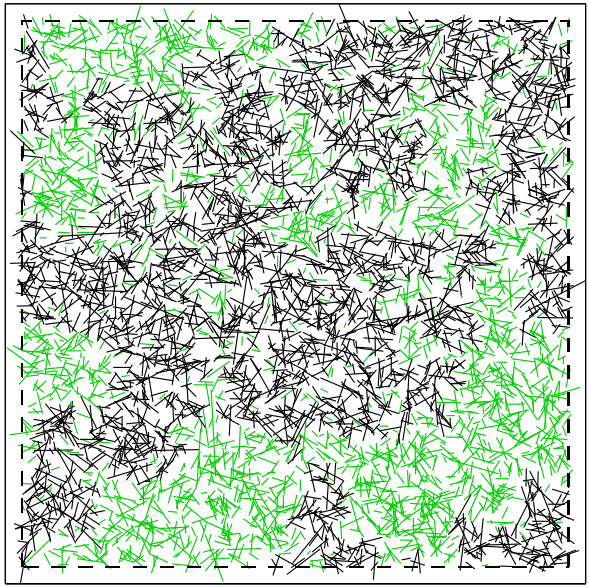}
  \caption{Example of a system under consideration with intermediate values of the parameters, $s = 0.5$, $\mathrm{SD} = 0.5$. $L=32$. The number density corresponds to the percolation threshold. The incipient wrapping cluster is highlighted.\label{fig:sample}}
\end{figure}

To check for any occurrences of wrapping clusters, we used the union--find algorithm~\cite{Newman2000PRL,Newman2001PRE} adopted to continuous percolation~\cite{Li2009PRE,Mertens2012PRE} and paired with the Machta algorithm~\cite{Machta1996PRE}. Sticks were added one by one onto the substrate until a  cluster wrapping around the torus in two directions had arisen. Figure~\ref{fig:sample} demonstrates an example of a system under consideration with intermediate values of the parameters ($s = 0.5$, $\mathrm{SD} = 0.5$) exactly at the percolation threshold (number of sticks is 4833, $n_c \approx 4.72$). The resulting critical number density was averaged over $10^5$ independent runs to obtain the probability  of percolation, $R_{N,L}$.

To obtain the probability $R(n,L)$ of percolation in the grand canonical ensemble, we convolved $R_{N,L}$ with the Poisson distribution~\cite{Li2009PRE,Mertens2012PRE}.
\begin{equation}\label{eq:convolution}
  R(n,L)= \sum_{N=0}^\infty \frac{\lambda^N \mathrm{e}^{-\lambda}}{N!} R_{N,L}.
\end{equation}
Note that
$$
\sum_{N=0}^\infty \frac{\lambda^N \mathrm{e}^{-\lambda}}{N!} = 1, \forall \lambda > 0.
$$
The weights in Eq.~\eqref{eq:convolution}
$w_N (\lambda)={\lambda^N}/{N!}$
can be calculated using the recurrent relations~\cite{Mertens2012PRE},
\begin{equation}\label{eq:weight1}
  w_{\bar{N}-k} =
  \begin{cases}
    1, & \mbox{for } k=0, \\
    \frac{\bar{N}- k + 1}{\lambda} w_{\bar{N} -k +1}, & \mbox{for } k=1,2,\dots,
  \end{cases}
\end{equation}
and
\begin{equation}\label{eq:weight2}
  w_{\bar{N}+k} =
  \begin{cases}
    1, & \mbox{for } k=0, \\
    \frac{\lambda}{\bar{N}+ k} w_{\bar{N} + k - 1}, & \mbox{for } k=1,2,\dots,
  \end{cases}
\end{equation}
herewith the relation
$
\sum_{N=0}^\infty w_N (\lambda) = \mathrm{e}^\lambda
$
should be borne in mind. Here, $\bar{N} = \lfloor \lambda \rfloor$.
Therefore, the convolution can be calculated as
\begin{equation}\label{eq:RNL}
R(n,L)= \sum_{N=0}^\infty w^\ast_N(\lambda) R_{N,L},
\end{equation}
where
\begin{equation}\label{eq:wstar}
  w^\ast_N(\lambda) = \frac{w_N(\lambda)}{\sum_{N=0}^\infty w_N(\lambda)}.
\end{equation}
Since
$$
\sum_{N=0}^\infty w_N (\lambda) = \mathrm{e}^\lambda \sum_{N=0}^\infty w^\ast_N (\lambda),
$$
$\mathrm{e}^{-\lambda}$ is absent in the master equation~\eqref{eq:RNL}.

\begin{figure}[!htb]
  \centering
  \includegraphics[width=\columnwidth]{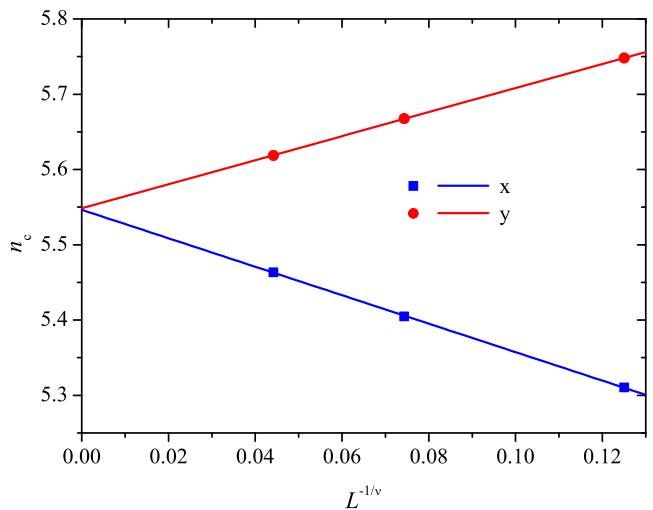}
  \caption{Example of scaling for $s=0.5$, $\mathrm{SD} = 0.5$.\label{fig:scaling}}
\end{figure}

Unfortunately, conformal field theory gives exact values for the wrapping probabilities at the
transition in the limit $L \to \infty$ only for isotropic systems~\cite{Pinson1994JSP,Newman2000PRL,Newman2001PRE}. The most effective method to estimate the percolation threshold~\cite{Newman2000PRL,Newman2001PRE,Li2009PRE,Mertens2012PRE} does not work when the system is anisotropic. This is the reason why a different, less efficient, approach was used in our study. For each particular value of $L$, the equation $R(n_c,L) = 0.5$ was solved numerically using bisection. Then, the scaling relation~\cite{Stauffer} was applied to find the percolation threshold in the thermodynamic limit
\begin{equation}\label{eq:sqaling}
  n_c(\infty) - n_c(L) \propto L^{-1/\nu}, \text{ where } \nu= 4/3.
\end{equation}
We used $L = 16, 32, 64$ to perform the scaling analysis; an additional size, $L=128$, was used for the set of parameters $s=0$, $\mathrm{SD} = 0$. Figure~\ref{fig:scaling} demonstrates an example of scaling for $s=0.5$, $\mathrm{SD} = 0.5$. All results presented in Section~\ref{sec:results} correspond to the thermodynamic limit.

\section{Results}\label{sec:results}
\begin{figure}[!htb]
  \centering
  \includegraphics[width=\columnwidth]{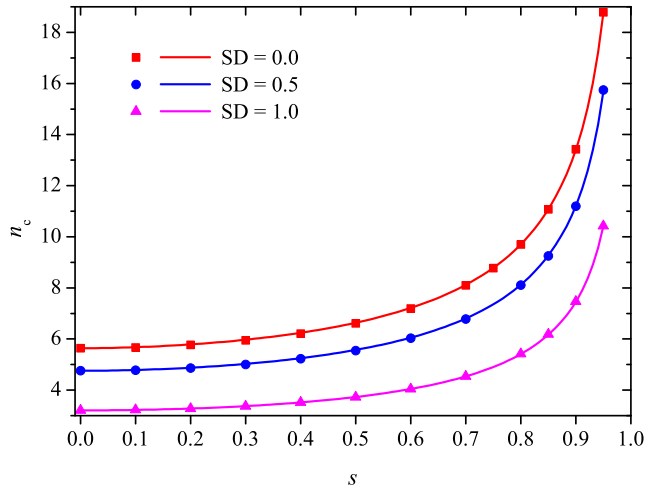}
  \caption{\label{fig:ncvss}Dependencies of the percolation threshold, $n_c$, on the order parameter, $s$, for different values of $\mathrm{SD}$. The curve corresponds to the least square fit~\eqref{eq:fit}.}
\end{figure}

Figure~\ref{fig:ncvss} demonstrates the dependencies of the percolation threshold, $n_c(s,\mathrm{SD})$, on the order parameter, $s$, for different values of $\mathrm{SD}$. For any value of $\mathrm{SD}$, the critical number density increases as the order parameter increases. The curves were fitted by
\begin{equation}\label{eq:fit}
  n_c(s,\mathrm{SD}) = \frac{n_c(0,\mathrm{SD})}{\sqrt{1 - s^\alpha}},
\end{equation}
where the fitting coefficient $\alpha$ depends on  $\mathrm{SD} $ (see Table~\ref{tab:params}.) From the nature of this case, $n_c(1,\mathrm{SD}) = \infty$, since percolation of parallel zero-width sticks is impossible for any finite value of the number density. The asymptotic behavior $n_c(s \to 1,0)$ corresponds to~\eqref{eq:asymp}.
 \begin{table}[!htb]
 \caption{Fitting parameter $\alpha$ in~\eqref{eq:fit}, and $n_c(0,\mathrm{SD})$ for different values of $\mathrm{SD}$.\label{tab:params}}
 \begin{ruledtabular}
 \begin{tabular}{cccc}
 $\mathrm{SD} $ & $n_c(0,\mathrm{SD})$& $\alpha$ & $R^2$ \\
  \hline
    0.0  & 5.63724(18) & 1.8449(26) & 0.99998\\
    0.5  & 4.756(3) & 1.880(5) & 0.99993\\
    1.0  & 3.21(1) & 1.9371(12) & 1\\
  \end{tabular}
 \end{ruledtabular}
 \end{table}

\begin{figure}[!htbp]
  \centering
  \includegraphics[width=\columnwidth]{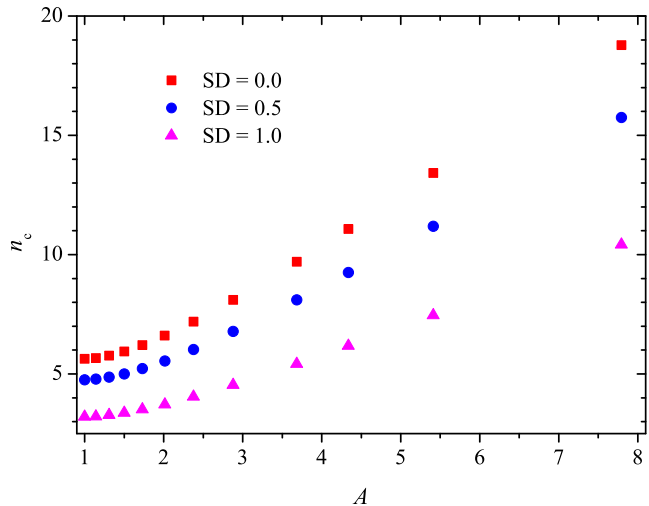}
  \caption{\label{fig:ncvsA}Dependencies of the percolation threshold, $n_c$, on the macroscopic anisotropy, $A$, for different values of $\mathrm{SD}$.}
\end{figure}

Figure~\ref{fig:ncvsA} shows the dependencies of the percolation threshold, $n_c$, on the macroscopic anisotropy, $A$, for different values of $\mathrm{SD}$. For the values of the macroscopic anisotropy $A \gtrapprox 3$, the dependencies look almost linear. However, any conclusion regarding their asymptotic behavior ($A \to \infty$) are not really possible.

\begin{figure}[!htbp]
  \centering
  \includegraphics[width=\columnwidth]{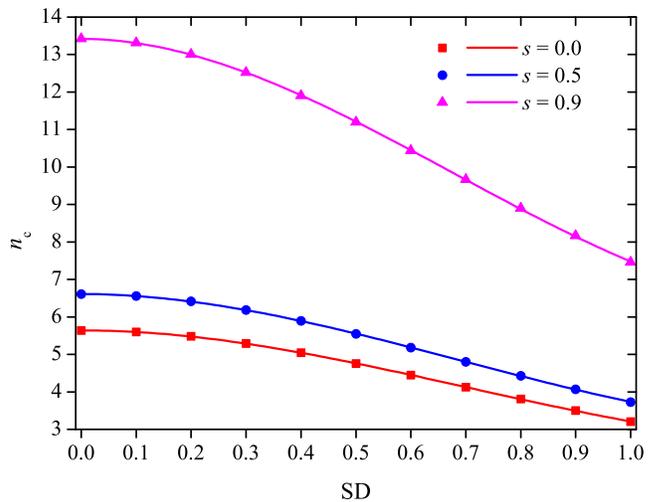}
  \caption{\label{fig:ncvssd}Dependencies of the percolation threshold, $n_c$, on the standard deviation of the length dispersity, $\mathrm{SD} $, for different values of the order parameter, $s$. The curve corresponds to the least square fit~\eqref{eq:fit1}.}
\end{figure}

Figure~\ref{fig:ncvssd} demonstrates the dependencies of the percolation threshold, $n_c(s,\mathrm{SD})$, on the $\mathrm{SD}$ for different values of the order parameter, $s$. For any value of $s$, the critical number density decreases as the order parameter increases. This behavior is not unexpected, since long sticks may appear when the length dispersity is large. These long sticks may assist the development of a percolating cluster even at low number densities. The curves were fitted by
\begin{equation}\label{eq:fit1}
  n_c(s, \mathrm{SD}) = n_c(s, 0) + a \cdot \mathrm{SD}^2 + b \cdot \mathrm{SD}^3,
\end{equation}
where the fitting parameters $a$ and $b$ depend on  $s$ (see Table~\ref{tab:params1}.)
 \begin{table}[!htb]
 \caption{Fitting parameters $a$ and $b$ in~\eqref{eq:fit1}, and $n_c(s,0)$ for different values of the order parameter,~$s$.\label{tab:params1}}
 \begin{ruledtabular}
 \begin{tabular}{ccccc}
 $s$ & $n_c(s,0)$ & $a$ & $b$ & $R^2$ \\
  \hline
   0.0 & 5.63724(18) & $-4.59(3)$  & 2.16(3) & 0.99995 \\
   0.5 & 6.6076(4)   & $-5.57(4)$  & 2.69(4) & 0.99996 \\
   0.9 & 13.422(4)   & $-11.72(9)$ & 5.78(9) & 0.99994 \\
  \end{tabular}
 \end{ruledtabular}
 \end{table}

\section{Conclusion}\label{sec:concl}
\begin{figure}[!h]
  \centering
  \includegraphics[width=\columnwidth]{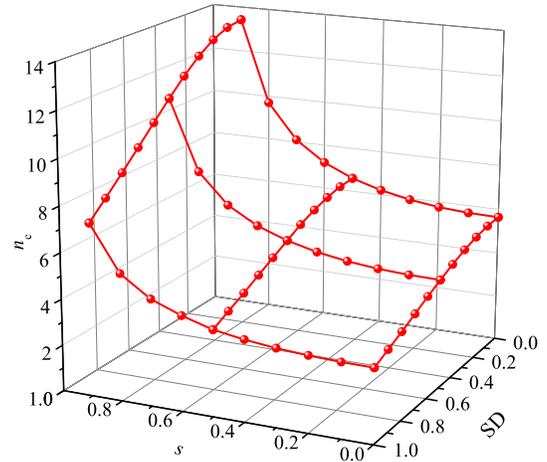}
  \caption{\label{fig:mesh}Dependence of the percolation threshold, $n_c$, on both the order parameter, $s$, and on the $\mathrm{SD}$.}
\end{figure}

By means of computer simulation and scaling analysis, we studied the percolation of zero-width sticks on a plane paying special attention to the cooperative effects of both the alignment of sticks and their length dispersity on the percolation threshold. The dependencies of the percolation threshold on the alignment of the rods and their length dispersity have been obtained in the thermodynamic limit. Figure~\ref{fig:mesh} demonstrates the dependence of the percolation threshold, $n_c$, on both the order parameter, $s$, and  on the $\mathrm{SD}$. The highest values of the percolation threshold correspond to highly anisotropic systems with equal-sized sticks while the lowest values correspond to isotropic systems with high length dispersity. The percolation threshold can be fitted as
$$
n_c(s,\mathrm{SD}) = \frac{n_c(0,0)  + a \cdot \mathrm{SD}^2 + b \cdot \mathrm{SD}^3 }{\sqrt{1 - s^\alpha}},
$$
where the coefficients $a$ and $b$ should be taken from the first row of Table~\ref{tab:params1} and $\alpha = 1.8449 + 0.0493 \cdot \mathrm{SD} + 0.04289 \cdot \mathrm{SD}^2$. An obvious drawback of our study is its consideration of only one particular kind of angular distribution and only one particular kind of length distribution. Nevertheless, we consider the chosen distributions as the most natural. For other kinds of physically reasonable distributions, similar behavior of the percolation threshold is expected.


\begin{acknowledgments}
We acknowledge the funding from the Ministry of Education and Science of the Russian Federation, Project No.~3.959.2017/4.6. 
\end{acknowledgments}

\bibliography{dispersity}

\end{document}